\documentclass[prb,twocolumn,showpacs,superscriptaddress,floatfix]{revtex4-1}
\usepackage{graphicx,amsfonts,amssymb,amsmath}
\usepackage[pagebackref=true,colorlinks,linkcolor=blue,citecolor=red]{hyperref}

\begin{document}
\title{Thin film growth by using random shape cluster deposition}

\author{ZH. Ebrahiminejad}
\affiliation{Department of Physics, Alzahra University,  Tehran, 19938, Iran}

\author{Seyed Farhad Masoudi}
\email{masoudi@kntu.ac.ir}
\affiliation{Department of Physics, K.N. Toosi University of Technology, P.O. Box 15875-4416, Tehran, Iran}

\author{R. S. Dariani}
\affiliation{Department of Physics, Alzahra University,  Tehran, 19938, Iran}

\author{Saeed S. Jahromi}
\affiliation{Department of Physics, K.N. Toosi University of Technology, P.O. Box 15875-4416, Tehran, Iran}

\begin{abstract}
The growth of a rough and porous thin surface by deposition of randomly shaped clusters with different sizes over an initially
flat linear substrate is simulated, using Monte Carlo technique. Unlike the ordinary Random Deposition, our approach
results in aggregation of clusters which produces a porous bulk with correlation along the
surface and the surface saturation occurs in long enough deposition times. The scaling exponents; the growth, roughness,
and dynamic exponents are calculated based on the time scale. Moreover, the porosity and its dependency to
the time and clusters size are also calculated. We also study the influence of clusters size on the scaling exponent,
as well as on the global porosity.

\end{abstract}
\maketitle 

\section{Introduction}
\label{intro}
Nowadays, the fabrication of the solid-state electronic devices
is vastly dependent on the deposition technology which is used in
their construction. The formation, growth and dynamics of
surfaces can well be regarded as the main factor in device
fabrication which has attracted strong attention, specially in
past few decades. Furthermore, the deposition of particles via
different procedures has immense importance not only in a broad
range of physics applications, but also in chemistry, biology,
and engineering. Until today, several analytical and
numerical approaches and approximation methods have been
developed to investigate the generic properties of various
surface growth processes. Since the nature of resulting surfaces
are random and far-from-equilibrium, they are called
non-equilibrium and rough surfaces \cite{Barabasi, Meakin}.

Regarding it as the simplest model of surface growth, Random
Deposition (RD) \cite{Barabasi} has a continuum growth equation which results in
exact determination of scaling exponents. In this model, each
particle is randomly dropped over a site of an initially flat
reference surface and deposited on the top of the selected
column. The growth of columns are independent of each other, and
the surface is uncorrelated. Ballistic deposition (BD) \cite{Vold} is another
method of surface growth which is a colloidal aggregation. In this model, a particle is
released from a randomly chosen position above the surface,
located at the distance larger than the maximum height of interface.
Since the different sites of the surface are dependent on the heights of neighboring sites,
a correlation develops along the surface. 
Hence, the height of the new particle will be equal to or larger than that of its neighbors.
The correlation length is the typical distance over which the heights "know about" each other which 
grows with time until it reaches the size of the system when the surface becomes saturated \cite{Barabasi}.
There are also other surface growth models such as
Random Deposition with Surface Relaxation (RDSR) \cite{Family1} and
nonlinear discrete models such as the Eden growth (ED) model \cite{Parisi},
restricted solid-on-solid (RSOS) model \cite{Kim}
and body-centered solid-on-solid (BCSOS) model \cite{Chin} have been proposed to study the
kinetic roughening of the growth processes which belong to different universality classes.

In order the simple surface growth equations to present more
complicated physical systems, theoretical efforts have been
carried out and different types of noise were added to the models
to adapt the possible variety of forces which might complicate
the dynamics of growth in each physical system. The continuum
linear growth model defined by the Edward-Wilkinson (EW) equation
is counted with the same universality class as that of the
discrete models, such as RDSR model \cite{Edwards}. The nonlinear
Kardar-Parisi-Zhang (KPZ) equation, which is a generalization of
the EW equation, explains the universality class which includes
the discrete growth models, such as the BD, RSOS, and BCSOS
models \cite{kpz}. In some approaches, the two or more different
deposition models are combined to investigate the time evolution
of roughness in real systems \cite{Wang, El-Nashar, Reis}. The
other works have proposed a surface growth model by the random
deposition of linear particles with different or identical sizes
in $(1+1)$ and $(2+1)$ dimensions. These particles are aggregated
on a flat surface with special deposition rules and a porous bulk
is generated \cite{Forgerini1, Forgerini2}.

Due to the daily development of nano-electronic technology, an
increasing demand is appeared to fabricate the nano-dimension
devices. Deposition of clusters instead of single particles due
to their large size, are a suitable choice for building the
desirable structures. Furthermore, as the clusters occupy more
than one unit site of the surface it is possible to generate
porous bulks which has plenty of applications in magnetic storage
\cite{Nielsch} , solar cells \cite{KARMHAG}, carbon nano-tubes
and other means \cite{Guangli, Che}. In the recent years, several
methods have also been proposed to simulate the growth of thin
films by deposition of clusters among which the Molecular
Dynamics (MD) simulations and Monte Carlo (MC) technique have
been frequently used. Unlike the MD model which is only
applicable to small systems, the MC outstandingly can be applied
to systems with large numbers of atoms \cite{Yanling, Biswas,
Kang, Mizuseki}.

In the following work, we are presenting a generalization of the particle deposition 
to enrich the physics of the simulation process and make it correspond more with the real 
deposition processes such as sputtering in which 
clusters of random shape and size are ejected from a solid target material due to bombardment 
of the target by energetic particles and deposited over the substrate. 
Although we do not deal with bombardment in our approach, we propose a method to produce
such random shape and size clusters, ranging from a single particle to a bunch of attached particles,
to make the simulation corresponds more with experimental reality. 
we are also interested to the study of the 
morphology of surfaces which are formed by adding such randomly shaped clusters to an initially 
flat linear substrate, based on MC simulations and random 
deposition model. 

In contrast to the ordinary RD in which there is no correlation along the surface
and the interface width increases infinitely without occurrence of saturation, our approach
results in aggregation of clusters which produces a porous bulk with correlation along the
surface. Moreover, the surface saturation is reached in long enough deposition times.
Consequently, the scaling exponents; the growth, roughness,
and dynamic exponents, are countable based on the time scale. In addition, the porosity and its dependency to
the time and clusters size is calculated. We have also
studied the influence of clusters size on the scaling exponent,
as well as on the global porosity.

The outline of the paper is as follows: In section \ref{model}, the model,
deposition rules, and the details involved in the Monte Carlo
simulations are described. The results of numerical simulations
for scaling exponents, interface width, and porosity as a function
of the time evolution and cluster's sizes are presented in section \ref{disscuss}.
The paper end with conclusion In section \ref{conclud}.
\section{ Model and Simulation Method for cluster deposition}
\label{model}
In the deposition process, the clusters with different shapes and size are
dropped randomly on an initially flat linear substrate. As deposition starts, the surface grows gradually.
The width of the surface which is called
the surface roughness $w$, is a consequence of the non-equilibrium growth
conditions under which the surface is formed
\begin{equation}
w(L,t)= \sqrt{\frac{1}{L}\sum_{i=1}^L[h(i,t)-\bar{h}(t)]},
\end{equation}
where $h(i ,t)$ is the height of the growing surface at
horizontal position $i$ and time $t$ and $L$ is the size of the substrate. $\overline{h}(t)$ is the
horizontally averaged surface height at time $t$
\begin{equation}
\overline{h}(t)=\frac{1}{L}{\sum_{i=1}^L}{h(i,t)},
\end{equation}
\begin{figure}
\centering{\includegraphics[width=\columnwidth]{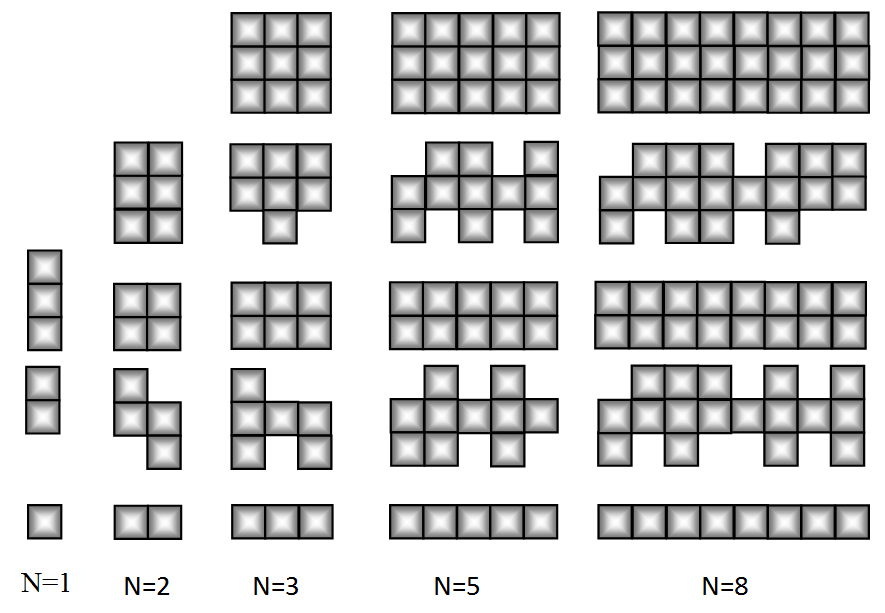}} \caption{Pictorial examples of clusters with different sizes and random shapes}
\label{fig1}
\end{figure}

The comprehensive scaling behavior of the rough surface is
studied with the Family-Vicsek scaling ansatz \cite{Family}. The
scaling relation implies that the interface width, increases as a
power of time, $w(t)\sim{t^{\beta}}$, which $\beta$ is the
roughness scaling exponent. The surface width saturates at a
value which increases as a power low of the system size,
$w(t)\sim{t^\alpha}$, where $\alpha$ is the growth scaling
exponent. These two exponents are related to each other by
another scaling exponent which is called the dynamic scaling
exponent; $z=\frac{\alpha}{\beta}$ \cite{Barabasi}.
\begin{equation}
w(L,t)={L^\alpha}f(\frac{t}{L^{z}}),
\end{equation}
The function $f(\frac{t}{L^{z}})$ is expected
to have an asymptotic form such that
\begin{equation}\label{w}
w(L,t)\sim\left\{\begin{array}{cc}
t^\beta, &  t\ll{L^z}.\\
L^\alpha, &  t\gg{L^z}.
\end{array}\right.
\end{equation}
In addition to roughness width, $w$, the correlation length is another parameter
which characterizes the rough surface by measuring the rate of
changes of roughness along the surface. As the RD
model has an uncorrelated nature, the correlation length is zero
and the roughness exponent is infinite. Therefor, the resulting
surface is not self-affine and the saturation will not be reach.
\begin{figure}
\centering{\includegraphics[width=\columnwidth]{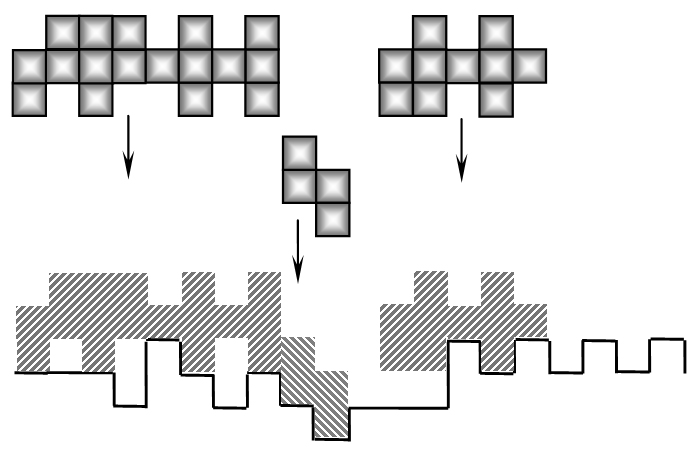}} \caption{Illustration of the 
growth formalism and geometry conditions of the cluster deposition}
\label{fig2}
\end{figure}

In the following, at first, we will describe the procedure of building and depositing the clusters
with different shape and size, using Monte Carlo simulation. Then, we show that during the surface
growth, three different behaviors appear as a function of time.
These behaviors at the initial and intermediate times,
are characterized by growth exponents $\beta_1$ and $\beta_2$,
respectively. At large times, the surface width grows slowly and the saturation will be reached. The morphology of
rough surface is then characterized by the value of the exponent $\alpha$. In addition, the variation of global porosity
with time and size of the bulk formed by depositing clusters are studied.

In our approach for constructing rough structures, instead of growing the surface by atom-atom aggregation on
the substrate, we propose a cluster growth procedure in which a random
number of single particles with a unit height are attached together in a way that
different shape clusters are formed.
Each cluster is a $3\times N$ linear box where $N$ is the cluster length
and ranges from one to an arbitrary random number. In other way, $N$ is the
maximum permitted size for the deposition.
The middle row of the cluster is fully filled with unit particles and the
top and bottom rows are occupied by random number of particles, $D$, $(0\leq D\leq N)$,
which their positions are quite random (Fig.~\ref{fig1}).

As the deposition is initiated, the clusters are dropped over random positions of a finite
one dimensional substrate. The middle point of the cluster
would land over the randomly selected position of the substrate.
Regarding the size of a cluster, its middle point is considered as the half of its length.
Thereafter, by checking the vacant sites around the landing point which depends on the geometric constraint of the
previous deposited layer (the previous MC step) the lowest possible place in which the
cluster can be fitted is determined the incident cluster permanently
sticks there. 

To clarify the growth rules and geometry conditions of the surface,
three situations for which the clusters might be added to the surface are illustrated in Fig.~\ref{fig2}.
The vacant locations across the surface which has no other chance of being filled on the next deposition step, produce
a porous structure in the bulk. The porosity is defined as the
proportion of the unoccupied part of the surface volume to the total
surface volume
\begin{equation}
P=\frac{V_e}{V_{tot}},
\end{equation}
where $V_{tot}=V_e+V_s$, which $V_s$ and $V_e$ are the volume of
the occupied and empty sites, respectively. As
the clusters are larger than unit site of surface, a porous
structure is produced such that its porosity is dependent on the
time evolution and size of the clusters \cite{poros1, poros2}. In
the next section, we focus on studying this dependency and scaling
exponent of the surface by cluster deposition.
\section{Discussion and Results}\label{disscuss}
We performed a MC simulations on a linear flat substrate with it's size
ranging from $L=2^7$ to $2^{11}$. The deposition process starts at time $t=0$
and the log-log curve of the time dependency of roughness for
different values of substrate and clusters size is evaluated. Furthermore, the
behavior of the surface width and scaling exponents are studied
for two cases of different and identical cluster's size.
The time evolution of the porosity and its dependency to the
cluster's size are also discussed in section \ref{poro}.
\subsection{Clusters with different size}\label{difsize}
Considering the clusters with different size, we investigate the time evolution of the
surface roughness and estimate the scaling exponents and determine the
universality class of the model. The MC simulations were performed 
for the substrates with different sizes as $L=128, 256, 512, 1024$ and
$2048$. The numerical results show that as time goes by, in spite of the RD model, the surface
roughness in our model presents three different behaves. This happens
due to the correlation among the columns which leads
to the lateral growth of the surface. The roughness behaviors correspond to the initial,
intermediate and long time scales. The unit of time is measured in Monte
Carlo steps (MCs).
\begin{figure}
\centering{\includegraphics[width=\columnwidth]{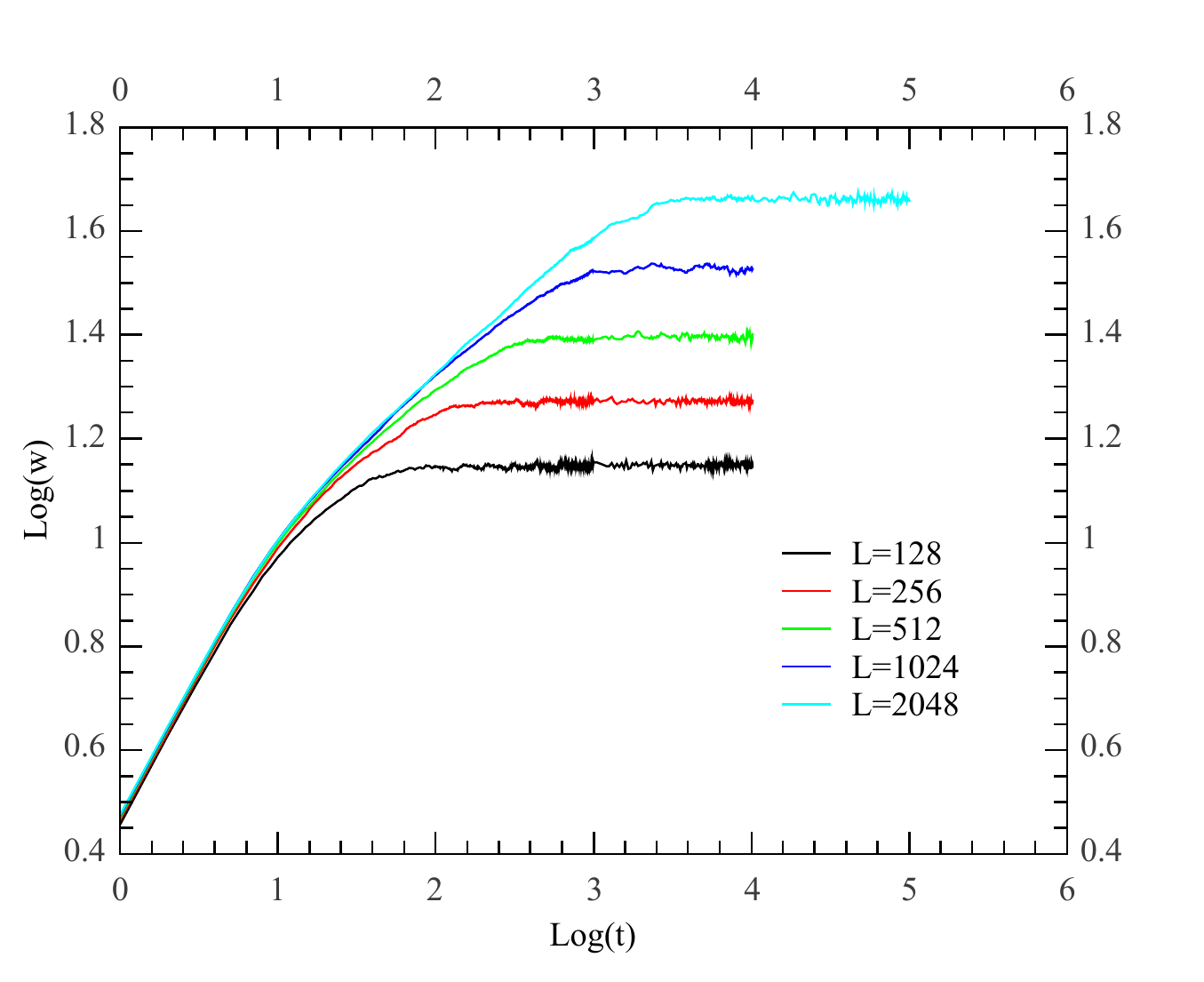}} \caption{Log-log plot of the roughness as a function of time for
deposition with different cluster shapes. From bottom to top the lattice sizes are L= 128, 256, 512, 1024, 2048.} 
\label{fig3}
\end{figure}

Fig.~\ref{fig3} demonstrate the log-log plot of the surface width as a function of time 
for the deposition of different size clusters on the substrate.
The results were obtained for the deposition of a mixture of clusters
with different sizes, ${3}\times{N}$, where N ranges from $1$ to $8$. In order to
get reliable results, the average sampling were performed over $10^3$ different
samples for $L=128-1024$ and $500$ for $L=2048$.
As it is illustrated in the figure, the behaviour of surface width at initial times
which corresponds to $\beta_1$, is close to $\frac{1}{2}$ for all substrate
sizes. This value is close to that of the random deposition of
particles with arbitrary correlation.
At the intermediate times, the lateral correlations are appeared
and the growth exponent $\beta_2$ shows a completely different behavior.
During this time, the height-height correlation develops and $\beta_2$
exhibits smaller value than $\beta_1$. At this times $\beta_2$
is approximately $1/4$ which in contrast to $\beta_1$, does
not belong to the RD universality. Although it is close to the results for the RDSR
model in (1+1) dimension which is described by EW equation, however it is possible that our model does not belong to any universality class.
It is due to the fact that the exponents have to be in exact agreement with that of EW so that we could decisively conclude that
they are in the same universality class. 
Eventually at long enough times, the  $\alpha$ exponent characterizes the roughness of
the saturated surface. The roughness width $w$ has a power law dependency to the $\alpha$ exponents
with respect to $L$ (i.e.,${w}\sim{L^{\alpha}}$).
In other words, surface grows in early
time regimes and saturates to a finite value, which depends on the
size L. The exponent $\alpha$ is estimated close to $1/2$. 
The scaling exponents were also measured for averaging over $1000$ independent MC runs
for each system size and the dynamic exponent $z$ obtained equal to $2.00$ from $\alpha$ and $\beta_2$.

\begin{figure}
\centering{\includegraphics[width=\columnwidth]{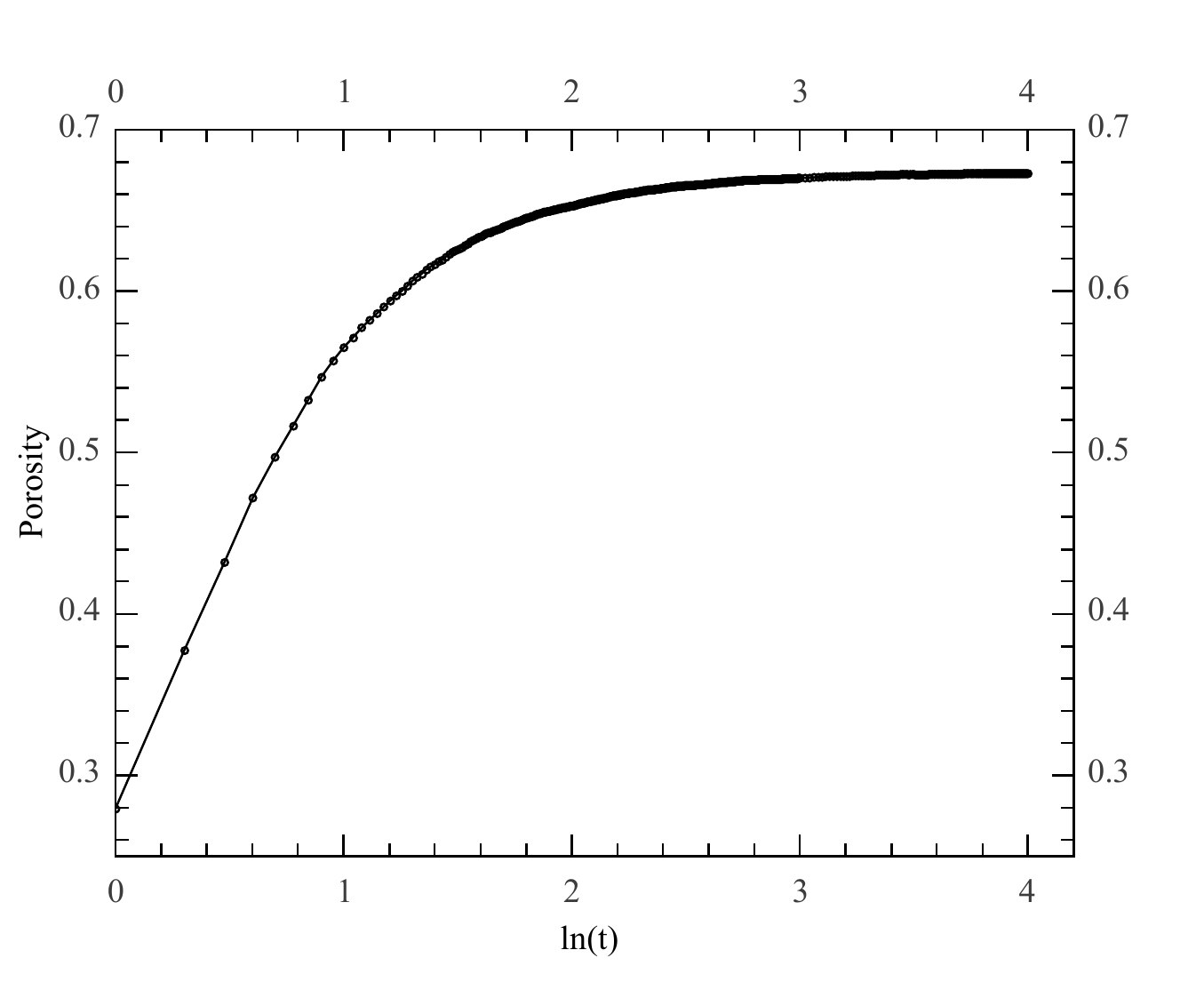}} \caption{Porosity as a function of time for
depositing clusters with different shape and size on a linear substrate of size L=1024}
\label{fig4} 
\end{figure}
\begin{figure}
\centering{\includegraphics[width=\columnwidth]{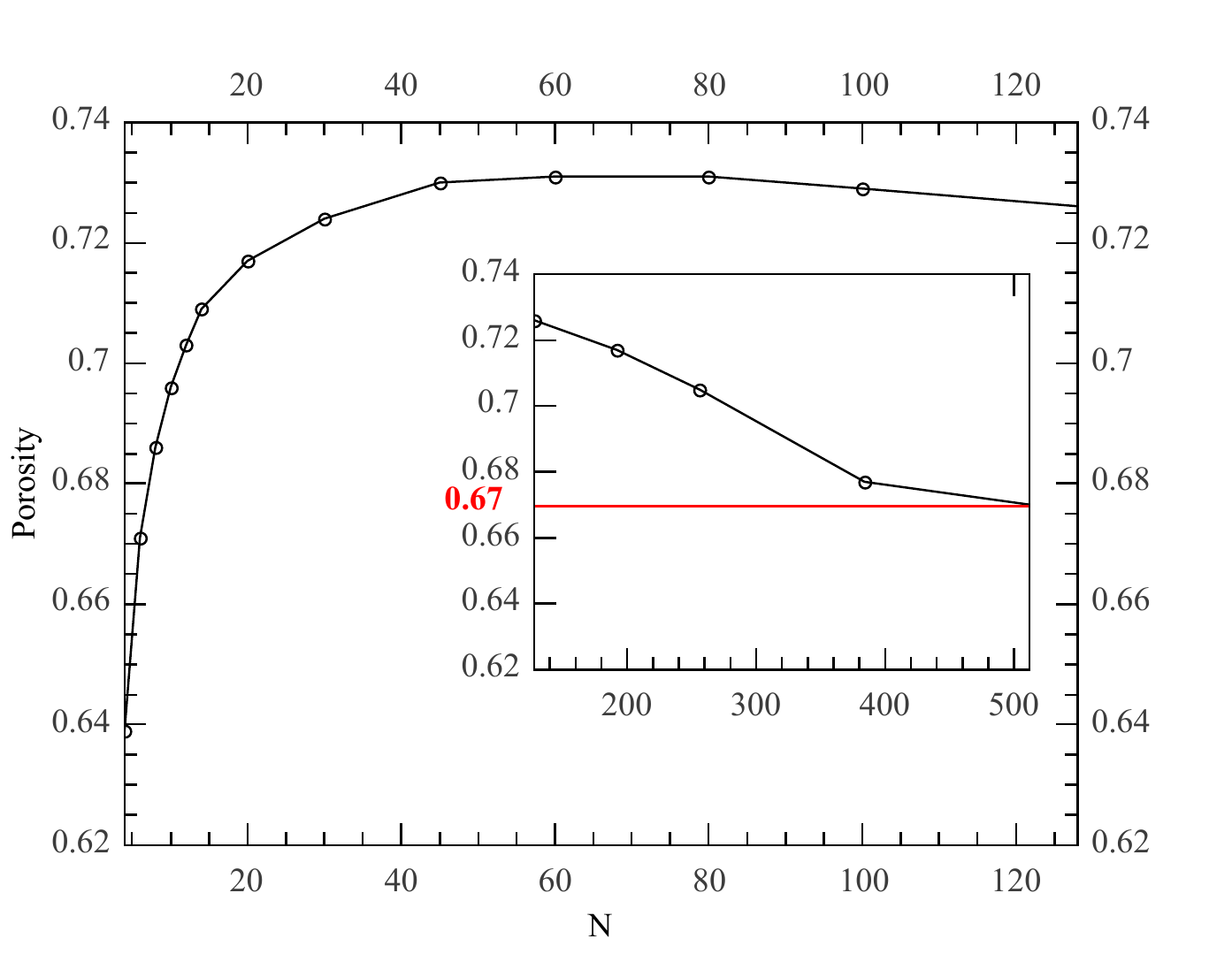}} \caption{Porosity versus cluster size for a linear substrate of
size L=1024} 
\label{fig5}
\end{figure}
\subsection{Clusters with identical size}\label{idsize}
Following our investigations,the MC simulations were performed for 7 different
cases for each of which the clusters of identical size($N$)
but with different shapes were deposited over the substrate (Table \ref{tab1}). 

The results show that the surface width does not exhibit a
noticeable dependency to the size of clusters, and the time
evolution of the roughness behaves in a similar manner to that of clusters of different sizes.
In this case also there are two different regimes at the initial and intermediate times which
correspond to the exponents $\beta_1$ and $\beta_2$,
respectively. The exponent $\alpha$ is also determined at 
the long enough times when the saturation is reached

Table \ref{tab1} demonstrate the results of each MC simulation for different cases of
deposition of identical size clusters. The scaling exponents
$\beta_1$, $\beta_2$, $\alpha$, and z for a substrate with L$=1024$ are presented in the table.
Although the clusters have identical sizes, the exponents behave the same as the previous approach with different size clusters.
Once again we claim that although the exponents are close to that of EW universality class but as they are not in exact agreement,
our model may not belong to any universality class.
For example in the case N=4, the scaling exponents explain the EW universality class. However, it might be accidental.
\begin{table}[!t]
 \begin{center}
 \caption{The growth, roughness and dynamics exponents for different clusters sizes}
 \label{tab1}
   \begin{ruledtabular}
   \begin{tabular}{lccccc}
    N      	    &$\beta_1$          &$\beta_2$       	  &$\alpha$        &$z$\\
    \hline
    2               &0.495$\pm$0.001    &0.293$\pm$0.001	  &0.52$\pm$0.02   &1.77$\pm$0.07  \\
    4		    &0.532$\pm$0.004	&0.256$\pm$0.001	  &0.50$\pm$0.03   &1.95$\pm$0.11  \\
    6               &0.545$\pm$0.005    &0.260$\pm$0.002	  &0.54$\pm$0.04   &2.07$\pm$0.15  \\
    8               &0.553$\pm$0.007	&0.268$\pm$0.001	  &0.55$\pm$0.03   &2.05$\pm$0.11  \\
    10              &0.557$\pm$0.006    &0.267$\pm$0.002	  &0.51$\pm$0.04   &1.91$\pm$0.15  \\
    12              &0.562$\pm$0.007    &0.289$\pm$0.003	  &0.58$\pm$0.05   &2.00$\pm$0.17  \\
    14              &0.580$\pm$0.004    &0.296$\pm$0.004	  &0.62$\pm$0.03   &2.09$\pm$0.10  \\
    \end{tabular}
  \end{ruledtabular}
 \end{center}
\end{table}

\subsection{Porosity}\label{poro}
\begin{figure}
\centering{\includegraphics[width=\columnwidth]{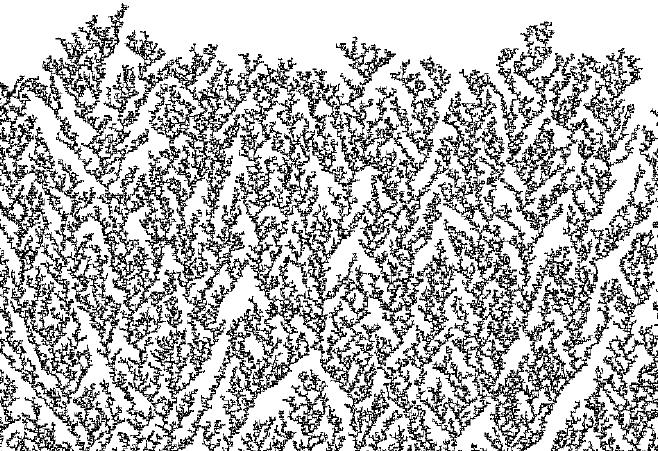}} 
\caption{Image of the top most pieces of the surface which is
obtained by our cluster deposition model.}
\label{fig6}
\end{figure}

Deposition of particles of size 2 or larger would lead to the
generation of porous structures \cite{Karmakar,Trojan,Caglioti}.
Our model is also based on the deposition of different size and
shape clusters and is expected to be porous. Fig.~\ref{fig6} illustrates a
piece of the porous surface which is obtained from our deposition
model. In the following, we investigate the porosity of our
surface as a function of time and cluster size.

Fig.~\ref{fig4} demonstrate the porosity of a surface of $L=1024$ with different size clusters
as a function of time. Unlike the roughness width, porosity saturates at the early stages of
deposition. The saturation occurs at $0.67$.
The dependency of global porosity of a surface of $L=1024$ to the cluster's size is depicted in Fig.~\ref{fig5}. 
The results show that as the cluster size is increased, the porosity is increased either. And after 
reaching a maximum value at $0.731$, the porosity is decreased and tends to $\frac{2}{3}$ in $N=\frac{L}{2}$.
One can explain this behavior by considering the variation of porosity in terms of the total height of the surface.
If the total number of deposited clusters and the maximum surface height are denoted by $M$ and $h$, respectively, the porosity 
can be approximated as:
\begin{equation} \label{poros}
P \simeq 1-\frac{2MN}{hL}
\end{equation}
Due to consideration of the overhangs from the topmost layer, the maximum value for porosity
is generally smaller than what is obtained from Eq.~\ref{poros}. Considering the $N=\frac{L}{2}$ case,
after depositing a cluster on one side of the substrate, the probability of depositing another cluster exactly adjacent
to it is very small ($\frac{1}{L}$). Hence, the second cluster is mostly probable to land over the previous one. As the height of each cluster in our
model is 3, the total height of the surface at the end of the growth process is $h=3M$. Inserting $h$ in Eq.~\ref{poros}, the porosity reads $P=\frac{2}{3}$.
For the case $N=\frac{L}{4}$, the second cluster sits in the vicinity of the first cluster with probability $\frac{3}{4}$ and over it with $\frac{1}{4}$.
The prosity is $\frac{2}{3}$ for the former and $\frac{6}{7}$ for the later and the total porosity is obtained as $P=0.71$.
Therefor, as the cluster size is decreased from $\frac{L}{2}$, the porosity is increased.
The increase of porosity continues until the blending and mixturing of clusters revers the process and the porosity starts to decrease.

Our investigations also shows that the porosity is independent of the linear
size of substrate.

\section{Conclusion}\label{conclud}
We have used the Monte Carlo simulations to study a
surface growth model with deposition of the mixture of clusters
with different sizes, $3 \times N$, where $N$ is the cluster length
and ranges randomly from one to a maximum value.

Due to deposition of
clusters with different sizes, a porous bulk is formed which
reaches to a saturation regimes at the initial Monte Carlo steps (faster than the roughness width).
The porosity depends on the cluster's size and increases as the
size of clusters is increased until it reaches a maximum  value, then decreases
and tend to the value $P=0.67$.

Our results show that surface width presents three different behavior as a function of time.
At initial times, the behavior is
close to an uncorrelated growth and the growth exponent is
$\beta_1\cong0.5$. At intermediate times, the surface roughness
grows more slowly with another growth exponent $\beta_2$ rangs $0.25-0.30$.
And eventually, at long times, surface reaches to a saturation regime
which is characterized by the roughness exponent $\alpha$ rangs $0.5-0.6$
and as a result, the dynamical critical exponent rangs $1.8-2.1$. The results certify that, 
the present deposition model does not belong to any universality class even though it's exponents are close to that of EW.

\vspace*{5mm}

\end{document}